\documentclass[12pt]{article}
\pdfoutput=1
\usepackage{latexsym, graphicx,cite}
\usepackage{amsmath}
\usepackage{amssymb}
\usepackage{amsthm}

\usepackage[top = 1. in,bottom = 0.93 in, left = 1 in, right=1 in]{geometry}

 \newcommand{\arXiv}[1]{\href{http://www.arXiv.org/abs/#1}{arXiv:#1}}
\usepackage[colorlinks=true, linkcolor=blue, bookmarks=true]{hyperref}

\makeatletter
\renewcommand\section{\@startsection {section}{1}{\z@}%
                  {-3.5ex \@plus -1ex \@minus -.2ex}
                  {2.3ex \@plus.2ex}%
                  {\normalfont\large\bfseries}}
\renewcommand\subsection{\@startsection{subsection}{2}{\z@}%
                   {-3.25ex\@plus -1ex \@minus -.2ex}%
                   {1.5ex \@plus .2ex}%
                   {\normalfont\bfseries}}
\makeatother

\newcommand{\beq}{\begin{equation}}
\newcommand{\eeq}{\end{equation}}
\newcommand{\ber}{\begin{array}}
\newcommand{\eer}{\end{array}}

\newcommand{\de}{\delta}

\newcommand{\eps}{s}

\newcommand{\ena}{\end{eqnarray}}
\newcommand{\beqa}{\begin{eqnarray}}
\newcommand{\eeqa}{\end{eqnarray}}
\newcommand{\bea}{\begin{eqnarray}}
\newcommand{\eea}{\end{eqnarray}}

\newcommand{\ad}{\hat\alpha^\dagger}
\newcommand{\al}{\hat\alpha}

\theoremstyle{remark}

\begin{document}
\begin{titlepage}
\begin{flushright}
\phantom{arXiv:yymm.nnnn}
\end{flushright}
\vspace{-5mm}
\begin{center}
{\huge\bf Quantum resonant systems,\vspace{2.5mm}\\ integrable and chaotic}\\
\vskip 15mm
{\large Oleg Evnin$^{a,b}$ and Worapat Piensuk$^{c}$}
\vskip 7mm
{\em $^a$ Department of Physics, Faculty of Science, Chulalongkorn University,
Bangkok, Thailand}
\vskip 3mm
{\em $^b$ Theoretische Natuurkunde, Vrije Universiteit Brussel and\\
The International Solvay Institutes, Brussels, Belgium}
\vskip 3mm
{\em $^c$ Theoretical and Computational Physics Group, Department of Physics,\\ 
and  Department of Mechanical Engineering,\\
King Mongkut's University of Technology Thonburi, Bangkok, Thailand}
\vskip 7mm
{\small\noindent {\tt oleg.evnin@gmail.com, piensuk.worapat@gmail.com }}
\vskip 20mm
\end{center}
\begin{center}
{\bf ABSTRACT}\vspace{3mm}
\end{center}
Resonant systems emerge as weakly nonlinear approximations to problems with highly resonant linearized perturbations. Examples include nonlinear Schr\"odinger equations in harmonic potentials and nonlinear dynamics in Anti-de Sitter spacetime. The classical dynamics within this class of systems can be very rich, ranging from fully integrable to chaotic as one changes the values of the mode coupling coefficients. Here, we initiate a study of quantum infinite-dimensional resonant systems, which are mathematically a highly special case of two-body interaction Hamiltonians (extensively researched in condensed matter, nuclear and high-energy physics). Despite the complexity of the corresponding classical dynamics, the quantum version turns out to be remarkably simple: the Hamiltonian is block-diagonal in the Fock basis, with all blocks of varying finite sizes. Being solvable in terms of diagonalizing finite numerical matrices, these systems are thus arguably the simplest interacting quantum field theories known to man. We demonstrate how to perform the diagonalization in practice, and study both numerical patterns emerging for the integrable cases, and the spectral statistics, which efficiently distinguishes the special integrable cases from generic (chaotic) points in the parameter space. We discuss a range of potential applications in view of the computational simplicity and dynamical richness of quantum resonant systems.

\vfill

\end{titlepage}


\section{Introduction}

Our focus in this article will be on studying properties of quantum systems with Hamiltonians
\beq
\hat H=\frac{1}2\sum_{n,m,k,l=0, \atop n+m=k+l}^\infty \hspace{-3mm} C_{nmkl} \ad_n\ad_m \al_k \al_l,
\label{ressyst}
\eeq
where $\alpha_n$ and $\ad_n$ with integer $n\ge 0$ are Hermitian-conjugate operators satisfying the standard creation-annihilation commutation relations
\beq
[\al_n,\ad_m]=\de_{nm},
\label{comm}
\eeq
while $C_{nmkl}$ are real numerical coefficients known as the {\it interaction coefficients} or mode-coupling coefficients. (As the Hamiltonian is Hermitian, one must evidently have $C_{nmkl}=C_{mnkl}=C_{klnm}$.) The resonance condition $n+m=k+l$ has been imposed on the summation. What drives our interest in this type of problems? We mention a few distinct reasons:
\begin{itemize}
\item Classical counterparts of resonant systems of the form (\ref{ressyst}) naturally emerge in weakly nonlinear analysis of PDEs whose linearized spectra of frequencies are highly resonant (differences of any two frequencies are integer in appropriate units). Concrete examples in the recent literature include studies of gravitational stability of Anti-de Sitter spacetimes \cite{FPU,CEV,BMR,islands,SNH,AdS4} motivated by the numerical observations of turbulent instabilities in \cite{BR} (a review can be found in \cite{rev2}), as well as related problems of small amplitude dynamics of nonlinear wave equations in Anti-de Sitter spacetimes \cite{BKS,CF,BHP1,BEL,BHP2}. These investigations further connect to a mathematically closely related problem motivated by completely different physics, namely, weakly coupled dynamics of Bose-Einstein condensates \cite{BDZ,cooper,fetter} in isotropic harmonic traps \cite{GHT,GT,BMP,BBCE,GGT,Fennell}.
\item A particularly simple system in the class (\ref{ressyst}), defined by $C_{nmkl}=1$ is Lax-integrable and has been studied under the name of the `cubic Szeg\H o equation' \cite{GG} with strong results on turbulent energy transfer to higher modes. Other systems, emerging from weakly nonlinear PDE analysis, display `partial integrability' in the sense of possessing classes of explicit analytic solutions \cite{CF,BEL,SNH,BBCE,AO}. At generic values of the interaction coefficients, one has an infinite-dimensional Hamiltonian system with cubic nonlinearities in the equations of motion, which is generally expected to display chaotic behaviors.
\item While the above examples explain how classical resonant systems emerge, in some cases, such as dynamics of Bose-Einstein condensates, these are fundamentally semiclassical approximations to quantum systems. The study of quantum resonant systems is thus physically motivated from such perspective as well. Irrespectively of, and in addition to this motivation, there is an extra reason to study quantum resonant system: it is the main message of this paper that these quantum systems are remarkably tractable, in many ways simpler than their classical counterparts, and they provide an interesting testbed for analyzing a given set of interaction coefficients $C_{nmkl}$, in particular in relation to possible integrability -- highly complementary to (often onerous) numerical simulations of the classical dynamics.
\item The expression (\ref{ressyst}), albeit without the resonance constraint $n+m=k+l$ (which can be viewed as a special choice of $C_{nmkl}$ in a general unrestricted sum), very commonly occurs in condensed matter and nuclear physics as the contribution to many-body Hamiltonians from 2-particle interactions. (The restriction of the sum to resonant quartets of particle states is specific to our current work, and it is precisely what opens the door for the explicit results we shall obtain below.) One particular line of work that has much in common with our current angle (including connections to random matrix theory) is studies of bosonic Gaussian embedded ensembles \cite{boson1,boson2,boson3,boson4}. For a textbook treatment, see \cite{kota}.
\item The fermionic counterparts of the embedded Gaussian ensembles we have just mentioned have been studied even more extensively as improvements of naive random matrix theory for large nuclei, starting with \cite{nuclear1,nuclear2} (textbook \cite{kota} covers these topics as well). These systems are obtained from (\ref{ressyst}) by replacing the bosonic creation-annihilation operators with their fermionic counterparts, removing the resonance condition, and treating the interaction coefficients $C_{nmkl}$ as random. This line of research has recently been given an extra boost through its conjectured connections to quantum black hole physics \cite{SYK1,SYK2,SYK3,SYK4,SYK5,SYK6,SYK7,SYK8}. Here, a different aspect of technical similarity with our investigations emerges: while in our case, the Hamiltonian will turn out to be block-diagonal with blocks of finite sizes due to the imposition of the resonant constraint, the fermionic version has a finite-dimensional space of states by construction. Both cases are thus reduced to diagonalizing large, finite-dimensional matrices. We particularly mention recent investigations of quartic tensor quantum mechanics \cite{tensor1, tensor2,tensor3}.
\item The remarkable simplicity of the quantum system (\ref{ressyst}) potentially makes it an excellent arena to study the usual array of questions appearing in the area of `quantum chaos.' In particular, the spectral statistics and its connections to integrability and chaos \cite{btint,BGS,haake} can be straightforwardly investigated and will form part of our treatment. Likewise, it should be possible to rather directly approach the questions of eigenstate thermalization \cite{ETH}, and emergence of chaotic classical trajectories from our quantum solutions. (A contemporary example of treating this type of questions for a system with a finite-dimensional space of state, in this case a spin chain, can be found in \cite{BCH}.)
\end{itemize}

The paper is organized as follows. We shall first describe in section 2 how classical reso\-nant systems emerge from weakly nonlinear analysis of PDEs and present a few particular examples of such systems that will be useful for our further study. In section 3, we shall describe the general block-diagonal structure of the Hamiltonian (\ref{ressyst}) that makes it possible to reduce finding its energy eigenvalues to diagonalizing finite-sized numerical matrices. In section 4, we shall analyze explicitly the 2-particle sector of such diagonalization, which can be done by elementary means. In section 5, we shall present the results of our numerical diagonalization in multi-particle sectors, with particular emphasis on the spectral statistics and its connections to integrability. Finally, we shall give a summary and outlook in the concluding section.

\section{The origins and properties of resonant systems}

One natural way to display the physics behind our quantum resonant systems (\ref{ressyst}) is to show how their classical counterparts emerge from weakly nonlinear analysis of physically relevant PDEs. These classical counterparts are simply Hamiltonian systems with the Hamiltonian
\beq
H=\frac{1}2\sum_{n,m,k,l=0, \atop n+m=k+l}^\infty \hspace{-3mm} C_{nmkl} \bar\alpha_n\bar\alpha_m \alpha_k \alpha_l,
\label{Hclass}
\eeq
with the bar denoting complex conjugation, and the symplectic form $i\sum_n d\bar\alpha_n\wedge d\alpha_n$, so that the equations of motion are 
\beq
i\,\frac{d\alpha_n}{dt}=\hspace{-3mm}\sum_{m,k,l=0, \atop n+m=k+l}^\infty \hspace{-3mm} C_{nmkl} \bar\alpha_m \alpha_k \alpha_l.
\label{basicequa}
\eeq

To illustrate how equations of the form \eqref{basicequa} naturally arise from weakly nonlinear PDEs, it is convenient to bring up the particularly simple case of the one-dimensional nonlinear Schr\"odinger equation in a harmonic trap,
\begin{equation}
i\,\frac{\partial \Psi}{\partial t}=\frac12\left(-\frac{\partial^2}{\partial x^2}+x^2\right)\Psi +g|\Psi|^2\Psi.
\label{NLS1d}
\end{equation}
One first observes that the linearized problem ($g=0$) is simply the harmonic oscillator Schr\"odinger equation. Its general solution is thus written as
\begin{equation}
\Psi=\sum_{n=0}^\infty \alpha_n \psi_n(x) e^{-iE_n t},\qquad E_n=n+\frac12,\qquad \frac12\left(-\frac{\partial^2}{\partial x^2}+x^2\right)\psi_n=E_n\psi_n,
\label{NLS1dlin}
\end{equation}
with constant $\alpha_n$. Small nonlinearities, corresponding to a small nonzero coupling $g$, induce slow drifts of $\alpha_n$, which cease being constant. One can simply re-express the evolution in terms of these slow drifts of $\alpha_n$ by substituting \eqref{NLS1dlin} into \eqref{NLS1d} and projecting on $\psi_n(x)$. This gives
\begin{equation} 
 i \,\frac{d\alpha_n}{dt}  = g\sum_{k,l,m=0}^\infty C_{nmkl} \,\bar \alpha_m \alpha_k  \alpha_l \,e^{i(E_n+E_m-E_k-E_l)t}, 
\label{ampltd}
\end{equation}
where $C_{nmkl}=\int dx \,\psi_n  \psi_m \psi_k \psi_l$. The last representation we have obtained has the following peculiar feature at small $g$: since $\dot\alpha_n$ is of order $g$, $\alpha_n$ vary very slowly, as anticipated, on time scale of the order of $1/g$. One the other hand, most of the terms on the right hand side, except for the \emph{resonant terms} satisfying $E_n+E_m-E_k-E_l\equiv n+m-k-l=0$, oscillate rapidly, on time scales of order one, because of the explicit time dependence in the exponential factor. It is natural to expect that, at small $g$, the effect of such oscillatory terms `averages out,' and they can be discarded from the equation without affecting the precision at leading order in $g$. It can in fact be proved as a mathematical theorem that solutions with and without non-resonant (oscillatory) terms remain close to each other on time scales of order $1/g$. (We recommend \cite{murdock} for a textbook treatment, while \cite{KM} provides a rigorous analysis aimed at mathematicians and specifically focusing on nonlinear Schr\"odinger equations.)  Discarding all terms on the right-hand side of (\ref{ampltd}), except for those satisfying the \emph{resonance condition} $n+m=k+l$, and renaming the `slow time' $gt$ into $t$ results in an evolution equation of the form \eqref{basicequa}. This procedure is known in the literature under many different names including the resonant approximation, the effective equation, time-averaging, etc.

Application of similar analysis to a number of related equations, some of them much more complex, results in resonant approximations of the form (\ref{basicequa}). This includes higher-dimensional nonlinear Schr\"odinger equations (equivalently called Gross-Pitaevskii equations in the context of Bose-Einstein condensation) \cite{GHT,GT,BMP,BBCE,GGT,Fennell}, nonlinear wave equations in Anti-de Sitter spacetime \cite{BKS,CF,BHP1,BEL,BHP2} and weakly nonlinear gravitational dynamics of Anti-de Sitter spacetime \cite{FPU,CEV,BMR,islands,SNH,AdS4}. The physics of the original system becomes completely encoded in the interaction coefficients $C_{nmkl}$ within the resonant approximation. These coefficients depend on the form of the linearized normal modes and nonlinearities, and may vary in complexity between elementary one-term expressions, and the extremely complicated formulas of the gravitational case \cite{CEV}, originating from the complexity of nonlinearities in Einstein's equations. In practical applications, the resonant system (\ref{basicequa}) is often written in a form that includes an $n$-dependent factor on the left-hand side as $i\omega_nd\alpha_n/dt=$... This factor can, however, be straightforwardly eliminated by redefining $\alpha_n=\tilde\alpha_n/\sqrt{\omega_n}$ and $C_{nmkl}=\sqrt{\omega_n\omega_m\omega_k\omega_l}\,\tilde C_{nmkl}$. In this sense, our specification of resonant sytems in the form (\ref{basicequa}) is completely general (and covers, for instance, all specific physically relevant cases we have mentioned).

The resonant systems (\ref{basicequa}) respect the conservation of the following two quantities, which will be of great importance for us:
\beq
N=\sum_{n=0}^\infty |\alpha_n|^2,\qquad M=\sum_{n=1}^\infty n |\alpha_n|^2.
\label{consrv}
\eeq
The first quantity $N$ will become the particle number in the quantum case, while the second quantity $M$ originates from the energy conservation within the linearized theory, if viewed in the context of deriving resonant approximations from nonlinear PDEs. From the standpoint of the Hamiltonian system (\ref{Hclass}), the conservation of these quantities is related to the two $U(1)$ symmetries given by transformations $\alpha_n\to e^{i\theta}\alpha_n$ and $\alpha_n\to e^{in\phi}\alpha_n$, where $\theta$ and $\phi$ are parameters. Note that the second symmetry requires the resonance constraint in the sum in (\ref{Hclass}), and this is what will enable our subsequent analysis of the quantum version. While the above two quantities are conserved for any values of the interaction coefficients $C_{nmkl}$, special choices may lead to extra conservation laws. See, for example, \cite{AO} for a very large class of resonant systems admitting an extra conserved complex bilinear, as well as other analytic structures.

In our treatment of quantum resonant systems, we shall focus on a few different specifications of $C_{nmkl}$, which will give us a panorama of available options:
\begin{itemize}
\item The `cubic Szeg\H o equation' of \cite{GG} corresponding to
\beq
C_{nmkl}^{(Sz)}=1.
\label{CSz}
\eeq 
This system is known to be classically Lax-integrable, and hence one expects abundant patterns in the solutions of the quantum version, as our analysis will in fact demonstrate.

\item Three specific choices belonging to the infinite-dimensional class of partially solvable resonant systems described in \cite{AO}. Namely, the resonant system of the maximally rotating sector of the conformally coupled cubic wave equation on a 3-sphere \cite{BEL},
\beq
C_{nmkl}^{(MRS)}=\frac1{1+(n+m+k+l)/2},
\label{CMRS}
\eeq
the resonant system of the rotationally symmetric truncation of the same equation, known as the `conformal flow' from \cite{CF},
\beq
C_{nmkl}^{(CF)}=\frac{1+\min(n,m,k,l)}{\sqrt{(1+n)(1+m)(1+k)(1+l)}},
\label{CCF}
\eeq
the resonant system of the lowest Landau level truncation of the resonant approximation to the Gross-Pitaevskii equation in a two-dimensional isotropic harmonic trap \cite{GHT,BBCE,GGT},
\beq
C_{nmkl}^{(LLL)}=\frac{((n+m+k+l)/2)!}{2^{n+m}\sqrt{n!m!k!l!}}.
\label{CLLL}
\eeq
It is not known at present whether these systems are integrable, but their classical versions display a number of special analytic patterns, and some explicit analytic solutions are known. We shall see that these systems show eigenvalue spacing distributions normally associated with integrable systems. They also display a number of common suggestive patterns in their energy eigenvalues, and since they represent three rather different points in the space of partially solvable resonant systems of \cite{AO}, one might hope that these patterns will be common to the other resonant systems in this class as well.
\item It is interesting to consider a small \emph{ad hoc} modification of (\ref{CCF}) given by
\beq
C_{nmkl}^{(modCF)}=\frac{1+(n+m+k+l)/4}{\sqrt{(1+n)(1+m)(1+k)(1+l)}}.
\label{CmodCF}
\eeq
We shall see that this seemingly innocuous modification of the formula radically changes the eigenvalue distribution of the quantum case. Our methodology is thus a sensitive indicator of systems that display special analytic structures.
\item It is also instructive to fill $C_{nmkl}$ with independent identically distributed random numbers drawn from some simple distribution. This will give a picture of what a generic system of the form (\ref{ressyst}) does. In practice, we have used a uniform distribution supported  on the interval $[0,1]$ to generate the random interaction coefficients. This choice reflects the observation that the interaction coefficients are positive in physically motivated resonant systems, though the precise details of this choice are unlikely to be qualitatively significant. One expects that the classical dynamics corresponding to such randomly constructed resonant systems is chaotic, which is supported by the eigenvalue distributions we shall observe in the quantum case.
\end{itemize}

It is worth mentioning (though it is not the primary motivation for our treatment) that, while we have presented weakly nonlinear analysis of PDEs as a way to motivate the emergence of systems of the form (\ref{basicequa}), which are then studied in their own right and quantized into (\ref{ressyst}), one could have also applied the weakly nonlinear analysis of our type directly to the Heisenberg equations of motion, say, of a quantum nonrelativistic bosonic field in a harmonic trap. This means that (\ref{ressyst}) is likely to provide accurate weakly nonlinear approximations to the evolution of Heisenberg operators of Bose-Einstein condensates (or related systems). One thus gets a potential analytic handle on physically relevant quantum interacting field theories of some special kinds. We shall nonetheless adopt a more modest view in the present treatment, and simply study (\ref{ressyst}) as is from a dynamical systems perspective.

\section{Block-diagonal structure of the Hamiltonian}

We now come to the key point of our treatment, which opens a pathway for explicit analysis of the Hamiltonian (\ref{ressyst}), namely its block-diagonal structure. The space of states of the system (\ref{ressyst}) can be constructed in terms of the Fock basis $|n_0,n_1,\ldots\rangle$ satisfying, for any $k$ and with $n_k$ being nonnegative integers,
\beq
\ad_k\al_k|n_0,n_1,\ldots\rangle=n_k|n_0,n_1,\ldots\rangle.
\eeq
The classical conservation laws (\ref{consrv}) translate in the quantum case to two operators
\beq
\hat N=\sum_{k=0}^\infty \ad_k \al_k,\qquad \hat M=\sum_{k=1}^\infty k \, \ad_k \al_k
\eeq
that commute with the Hamiltonian (\ref{ressyst}). Since the Fock basis vectors are eigenvectors of these two operators
\beq
\hat N |n_0,n_1,\ldots\rangle=\left(\sum_{k=0}^\infty n_k\right)|n_0,n_1,\ldots\rangle,\qquad \hat M |n_0,n_1,\ldots\rangle=\left(\sum_{k=1}^\infty k\,n_k\right)|n_0,n_1,\ldots\rangle,
\eeq
it is guaranteed that the Hamiltonian has vanishing matrix elements between $|n_0,n_1,\ldots\rangle$ and $|n'_0,n'_1,\ldots\rangle$ unless the two sets of occupation numbers $\{n_k\}$ and $\{n'_k\}$ correspond to the same values of 
\beq
N=n_0+\sum_{k=1}^\infty n_k\qquad\mbox{and}\qquad M=\sum_{k=1}^\infty k\,n_k.
\label{conspart}
\eeq
The Hamiltonian is thus block-diagonal in the Fock basis, with blocks labelled by two nonnegative integers $(N,M)$. Note that all blocks are of varying finite sizes, since $n_k$ are nonnegative integers and hence there is only a finite number of different ways to satisfy (\ref{conspart}) for each given $(N,M)$. We have thus reduced the problem of solving the (very special) interacting quantum field theory (\ref{ressyst}) to diagonalizing finite-sized numerical matrices!

What about the sizes of the blocks? For each given $M$, (\ref{conspart}) corresponds to an integer partition of $M$ in which each integer number $k\ge 1$ is present $n_k$ times. Then the expression for $N$ says, remembering that $n_0$ is a nonnegative integer, that the number of these parts, given by $\sum_{k=1}^\infty n_k=N-n_0$, is less than or equal to $N$. Thus, the number of solutions to (\ref{conspart}) is simply the number of partitions of $M$ into at most $N$ parts, a well-known number-theoretical function usually denoted as $p_N(M)$. A practical summary of the properties of this function can be found in \cite{part}, while a standard textbook treatment of the subject is in \cite{andrews}.  For each $(N,M)$-block one then has to diagonalize a $p_N(M)\times p_N(M)$ matrix whose entries are real numbers made from the interaction coefficients $C_{nmkl}$. Such matrices are typically rather sparse, since the Hamiltonian has two annihilation and two creation operators and hence can only change $n_k$ in at most four positions and in a coordinated way (because of the resonant constraint). Thus, matrix elements between vectors that differ by more than that are guaranteed to be zero. We do not know, however, an explicit way to characterize this sparseness in terms of, say, having a limited number of nonzero diagonals.

The function $p_N(M)$ enjoys relatively slow growth at large $N$ and $M$ for an object of combinatorial nature. For fixed $N$ and large $M$, the known asymptotics \cite{part} is
\beq
p_N(M)\sim\frac{M^{N-1}}{N!(N-1)!},
\eeq
thus the growth is polynomial. Another relevant regime is $N=M$, in which case $p_M(M)$ is simply the total number of partitions $p(M)$, since there cannot be any integer partitions of $M$ into more than $M$ parts.
The large $M$ asymptotics of the total number of partitions is given by the famed Hardy-Ramanujan-Rademacher formula \cite{andrews} whose leading term is
\beq
p(M)\sim \frac{1}{4M\sqrt{3}}\exp\left(\pi\sqrt{\frac{2M}{3}}\right).
\eeq
The  growth is thus faster than polynomial, but still much slower than, say, $e^M$.

We note that partitions, and therefore the state vectors within our $(N,M)$-blocks can be generated recursively with respect to $N$ and $M$. Evidently, each partition of $M$ into at most $N$ parts is either a partition of $M$ into at most $N-1$ parts, or it is a partition of $M$ into exactly $N$ parts, the latter being in one-to-one correspondence with partitions of $M-N$ into at most $N$ parts (the correspondence is obtained by subtracting 1 from each element of the original partition into exactly $N$ parts, or if the element is 1, erasing it). This gives the standard recursion relation for $p_N(M)$,
\beq
p_N(M)=p_{N-1}(M)+p_N(M-N).
\eeq
It should be possible to extend this recursive definition to the vectors of states within the $(N,M)$-blocks, and their matrix elements, which is likely to be of use in future analytic considerations of quantum resonant systems.

The block-diagonal structure of (\ref{ressyst}) is highly amusing from the standpoint of imagining the semiclassical limit of the corresponding dynamics. Indeed, when dealing with systems characterized by finite-dimensional spaces of states, such as spin chains, one usually thinks that enlargement of the space of states (for instance, through enlargement of the magnitude of individual spins of the chain) must accompany taking semiclassical limits. Indeed, spaces of states emerging from quantizing conventional classical mechanical systems are automatically infinite-dimensional. In our case, the total space of states is of course infinite-dimensional, but the quantum evolution separates it into finite-size blocks. The classical dynamics must emerge from cross-talk between these finite-sized quantum blocks in a way that involves infinitely many states. An explicit analysis of this issue would likely be very curious, especially in view of the sophisticated and broadly studied classical dynamics of (\ref{basicequa}), though we shall not pursue this line of thought further at this point.

A note is in order on the  normalization of energies we adopt. By changing the energy scale  in (\ref{ressyst}), one can always set $C_{0000}=1$.  We shall be assuming this normalization wherever we talk about absolute magnitudes of energies. (Many of our results, however, refer to ratios of energies, which makes this normalization choice irrelevant.)

\section{Two-particle spectra}

We shall first discuss the blocks with $N=2$. While we shall see below that, in terms of displaying suggestive patterns, blocks with large $N$ and large $M$ are more attractive, the two-particle case has the advantage that much can be understood fully analytically. We shall resort to computer algebra in the next section to look for patterns in more complicated multi-particle cases.

For convenience, we shall specify $M=2m+1$ which simplifies the algebra. The case of even $M$ is, however, essentially identical. The relevant Fock vectors are very simple in this case. There are $m+1$ of them, and they will be labelled by $I=0,1,\ldots,m$. Each such vector $v_I$ is defined to have $n_I=1$ and $n_{M-I}=1$, with all other $n_k$ being zero. The matrix elements of the Hamiltonian between such states are easily computed in terms of the interaction coefficients as
\beq
H_{IJ}\equiv\langle v_I|\hat H|v_J\rangle=2C_{I,M-I,J,M-J}.
\label{H2prt}
\eeq
What remains is to diagonalize this $(m+1)\times(m+1)$ matrix for the cases of interest.

\subsection{The integrable Szeg\H o case}

For the interaction coefficients corresponding to the `cubic Szeg\H o equation' given by (\ref{CSz}), one simply has
\beq
H_{IJ}=2.
\eeq
Then, evidently, there are $m$ eigenvalues equal 0 for this matrix with identical entries, and a single eigenvalue equal $2(m+1)=M+1$. The high degeneracy at zero energy is a telltale sign that we are not dealing with a generic system. Many further special patterns will emerge in our subsequent analysis of the integrable Szeg\H o case.

\subsection{Partially integrable cases}

We now proceed with the three representative cases (\ref{CMRS}, \ref{CCF}, \ref{CLLL}) we have chosen out of the infinite class of resonant systems with special analytic properties presented in \cite{AO}:
\begin{itemize}
\item For (\ref{CMRS}), one has
\beq
H_{IJ}=\frac2{1+M}.
\eeq
Thus again, we get a matrix with identical entries, and the analysis is the same as above. There are $m$ zero eigenvalues, and one eigenvalue equal 1.

\item For (\ref{CCF}), we get
\beq
H_{IJ}=\frac{2(1+\min(I,J))}{\sqrt{(1+I)(1+M-I)}\sqrt{(1+J)(1+M-J)}}.
\label{H2CF}
\eeq
(This matrix can be related to the Lehmer matrix $\min(k,l)/\max(k,l)$, where $k$ and $l$ take values starting from 1, and not from 0 as in our labelling of rows and columns.) We first establish the qualitative structure of the spectrum  of (\ref{H2CF}).  From the proof of Lemma 3.1 in \cite{BHP1}, for any real $x_I$, one has
\beq
\sum_{K=0}^m (1+K)(1+M-K)x_K^2-2\sum_{I,J=0}^m(1+\min(I,J))x_Ix_J=2\sum_{K=0}^{m-1}\sum_{J=0}^m (1+K) (x_K-x_J)^2\ge 0.\nonumber
\eeq
Substituting $x_K=y_K/\sqrt{(1+K)(1+M-K)}$, we get
\beq
\sum_{K=0}^m y_K^2-\sum_{I,J=0}^m H_{IJ}\,y_I\,y_J=2\sum_{K=0}^{m-1}\sum_{J=0}^m (1+K) (x_K-x_J)^2\ge 0.
\eeq
Then, evidently, eigenvalues of $H_{IJ}$ cannot be greater than 1, while the eigenvalue 1 is attained for the eigenvector $y_K=\sqrt{(1+K)(1+M-K)}$. Similarly, one can write
\beq
\sum_{I,J=0}^m (1+\min(I,J))x_Ix_J=\sum_{I=0}^m\left(\sum_{J=I}^m x_J\right)^2\ge 0.
\eeq
Re-expressing in terms of $y_K$ like above, we conclude that all eigenvalues of $H_{IJ}$ are positive. To summarize, all $m+1$ eigenvalues of (\ref{H2CF}) lie between 0 and the maximal eigenvalue, which always equals 1. In fact, a closed form expression for the energy levels can be found,\footnote{We thank Piotr Bizo\'n for pointing out this formula to us. The matrix we are diagonalizing has previously appeared in studies of classical stability of the same model \cite{BHP1}.} with $I=0,1,\ldots,m$ (arranged, somewhat unconventionally, from the highest to the lowest energy):
\beq
E_I=\frac1{(I+1)(2I+1)}.
\eeq
\item For (\ref{CLLL}),
\beq
H_{IJ}=\frac1{2^{M-1}}\sqrt{M\choose I}\sqrt{M\choose J},
\eeq
with the standard notation for the binomial coefficients. Hence, if $y_J=\sqrt{M\choose J}$, then $\sum_{J=0}^m H_{IJ}\,y_J=y_I$. This gives a single eigenvector with eigenvalue 1. On the other hand, if  $y_J=\left[M\choose J\right]^{-1/2}x_J$ with $\sum_J x_J=0$, then $\sum_{J=0}^m H_{IJ}\,y_J=0$, which gives $m$ vectors with eigenvalues 0.

\end{itemize}

To summarize, the partially solvable examples possess, for any $M$, a spectrum entirely contained between 0 and 1, with exactly one eigenvector at the maximal eigenvalue 1. We shall see an extension of this pattern to multiparticle states, where we will have to rely on numerical experiments.

\subsection{Generic systems}

A `generic' resonant system may be understood as drawing $C_{nmkl}$ as independent identically distributed variables from some ensemble, subject to the index permutation symmetry. If this is done, $H_{IJ}$ defined by (\ref{H2prt}) is a real symmetric matrix whose entries are independently identically distributed, which is a completely standard formulation in random matrix theory \cite{mehta,GMW}. Then all the standard results or random matrix theory will apply, including the Wigner semicircle distribution for the eigenvalues, and the level spacing distribution of the Gaussian orthogonal ensemble. Note that many random matrix properties hold in particular realizations drawn from the ensemble, rather than only on average. For example, appropriately smoothed eigenvalue distributions of very large fixed random matrices (large $M$) drawn from such ensembles will approach the semi-circle law. This can be seen as an extreme version of the BGS conjecture \cite{BGS}, which states that level spacing statistics for (strongly) chaotic systems generically matches random matrix results. For the particular very simple case we are considering here (2-particle states of resonant systems with generic interaction coefficients), the statement is much stronger, since all aspects of statistics (not just level spacing) are literally identical to random matrices. We shall explore these questions in more detail for multiparticle states in the next section.

\section{Numerical experiments and spectral statistics}

We now proceed with the main technical part of our treatment, which is the analysis of energy eigenvalues of the Hamiltonian (\ref{ressyst}) within large $(N,M)$-blocks. While the abundant patterns we shall see make one suspect that an analytic treatment should be possible at least in some cases, at this time, we shall resort to numerical diagonalization.

There are many different relations between $N$ and $M$ one could consider as the size of the block, given by $p_N(M)$, increases. For example, $N$ can be kept fixed and small and $M$ taken large. This is a straightforward generalization of the previous section with $N=2$. Or $N$ and $M$ can be made large simultaneously. While we have performed numerical experiments at a number of different assignments of $N$ and $M$, the plots will be given for the case $N=M$ where particularly neat and simple patterns are observed, while we shall comment on the other cases briefly. Some of the patterns, such as the empirical formulas for maximal eigenvalues, are valid for any $N$ and $M$, according to our numerical observations.

\subsection{Numerical implementation}

To diagonalize the Hamiltonian (\ref{ressyst}) within a given $(N,M)$-block, one must first construct all the $p_N(M)$ normalized state vectors generated by the action of creation operators on the vacuum
\beq
\frac1{\sqrt{n_0!\,n_1!\cdots n_M!}}\left(\ad_0\right)^{n_0}\left(\ad_1\right)^{n_1}\cdots \left(\ad_M\right)^{n_M}|0\rangle
\label{statevec}
\eeq
with $n_k$ satisfying (\ref{conspart}).
(Evidently, $n_k=0$ for $k>M$ for any state belonging to an $(N,M)$-block.) Thereafter, one simply computes the matrix elements of (\ref{ressyst}) between such states, which can be implemented (depending on the desired efficiency) by repeated application of (\ref{comm}) and $\al_i|0\rangle=0$, or 
\beq
\langle 0|\left(\al_0\right)^{q_0}\left(\al_1\right)^{q_1}\cdots\left(\al_M\right)^{q_M}\left(\ad_0\right)^{r_0}\left(\ad_1\right)^{r_1}\cdots \left(\ad_M\right)^{r_M}|0\rangle=q_0!\,q_1!\cdots q_M! \,\de_{q_0r_0}\de_{q_1r_1}\cdots \de_{q_Mr_M}.\nonumber
\eeq
The result is an expression for the matrix elements of (\ref{ressyst}) in terms of the interaction coefficients $C_{nmkl}$. Once the numerical values of the coefficients have been substituted, one is left with an explicit $p_N(M)\times p_N(M)$ matrix with real entries, which can be diagonalized numerically.

For the evaluation of the matrix elements, we have found FORM particulatly suitable. FORM is a script-based computer algebra system \cite{form} designed for processing very large polynomial-type expressions and capable of implementing sophisticated pattern-matching and substitutions. After the matrix elements within a given $(N,M)$-block had been constructed with FORM, the resulting finite-sized numerical matrices were diagonalized using Sage \cite{sage}, on open-source computer algebra platform. (This diagonalization is nearly instantaneous on an ordinary personal computer for matrices whose size is of the order of a few thousand.)

To effectively generate all state vectors within a given $(N,M)$-block, one can resort to the following trick.
Introduce two parameters $\eta$ and $\xi$ and consider the following `master state'
\beq
\prod_{k=0}^M\left(\sum_{n=0}^{\min\left(\left\lfloor{M/k}\right\rfloor,N\right)}\frac1{\sqrt{n!}}\,\eta^n\,\xi^{kn}\left(\ad_k\right)^{n}\right)|0\rangle.
\eeq
(This sum is, of course, closely related to generating functions for partition numbers.) Evidently, the coefficient of $\eta^N\xi^M$ of this expression is simply a sum of all vectors of  the form (\ref{statevec}) belonging to the given $(N,M)$-block. In the context of FORM manipulations, it is convenient to generate this sum first, separate it into individual terms, and then treat them as state vectors used for the evaluation of the matrix elements.

We shall be talking below, in particular, about distributions of eigenvalues, or normalized distances between eigenvalues, in fixed large matrices. This, of course, requires some form of smoothing. The typical procedure we have in mind is to select an integer number $\Delta$ of magnitude comparable to $\sqrt{p_N(M)}$ for each $(N,M)$-block that has $p_N(M)$ eigenvalues, and plot various histograms with bins of size $\Delta$. (We normalize all of our histograms as probability distributions, so that the total area under each distribution curve is 1.)  Since $\Delta/p_N(M)\to 0$ as the block size given by $p_M(N)$ grows, while each bin still typically contains a large number of data points, one might hope that smooth limits of such distributions are attained as the block size increases. Our further numerical analysis suggests that this is indeed the case.

\subsection{Eigenvalue patterns for integrable cases}
\label{intpttrn}

We have performed numerical diagonalization for a number of different values of $N$ and $M$ following the algorithm outlined above. As one would expect, peculiar patterns emerge for the integrable Szeg\H o case, and also for the partially solvable cases of \cite{AO}. We summarize these patterns below.

\begin{figure}[t]
\hspace{4cm}\includegraphics[scale=0.4]{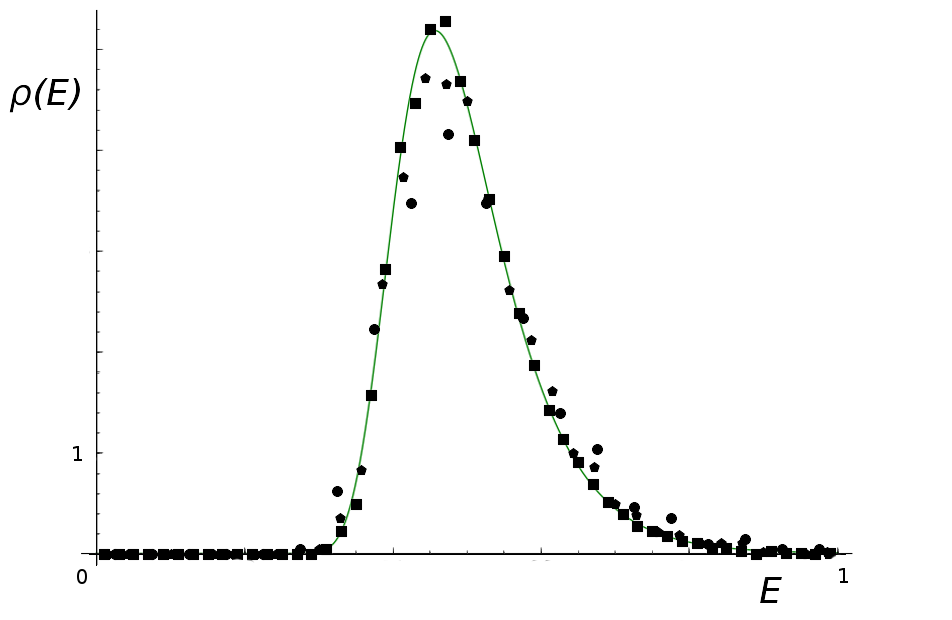}  
\caption{Distributions of normalized eigenvalues for the integrable Szeg\H o case. The normalized energy $E$ is measured in units of the largest eigenvalue in each block. Three cases are plotted: $N=M=18$ corresponding to a $385\times 385$ matrix (circles), $N=M=23$ corresponding to a $1255\times 1255$ matrix (pentagons) and $N=M=27$ corresponding to a $3010\times 3010$ matrix (squares). All points cluster around the same bell-shaped limiting curve, which can be identified as the (plotted) Gumbel distribution (\ref{gumbl}) after fitting $\mu$ and $\beta$. }
\label{eigSz}
\end{figure}

For the integrable Szeg\H o case (\ref{CSz}):
\begin{itemize}
\item All energy eigenvalues are \emph{integer} and non-negative, with the maximal eigenvalue within each $(N,M)$-block given by
\beq
E_{max}^{(Sz)}=\frac{(N-1)(N+2M)}2.
\eeq
\item The eigenvalue 0 acquires a very large multiplicity if $M\gg N$. Within such blocks, a substantial fraction of all eigenvalues are 0.
\item As the size of the blocks increases, the distribution of $E/ E_{max}^{(Sz)}$ (eigenstate energy normalized with respect to the largest energy eigenvalue) appears to converge to simple-looking bell-shaped curves, as long as $N$ and $M$ are both large. In Fig.~\ref{eigSz}, this is demonstrated for $N=M$, where distributions for three such blocks are plotted and clearly cluster along the same curve, which can be identified as the Gumbel distribution \cite{extreme}
\beq
\rho(E)=\frac{d}{dE} \exp\left[-e^{-(E-\mu)/\beta}\right].
\label{gumbl}
\eeq
\end{itemize}\vspace{3mm}
\begin{figure}[t]
\hspace{0.3cm}\includegraphics[scale=0.35]{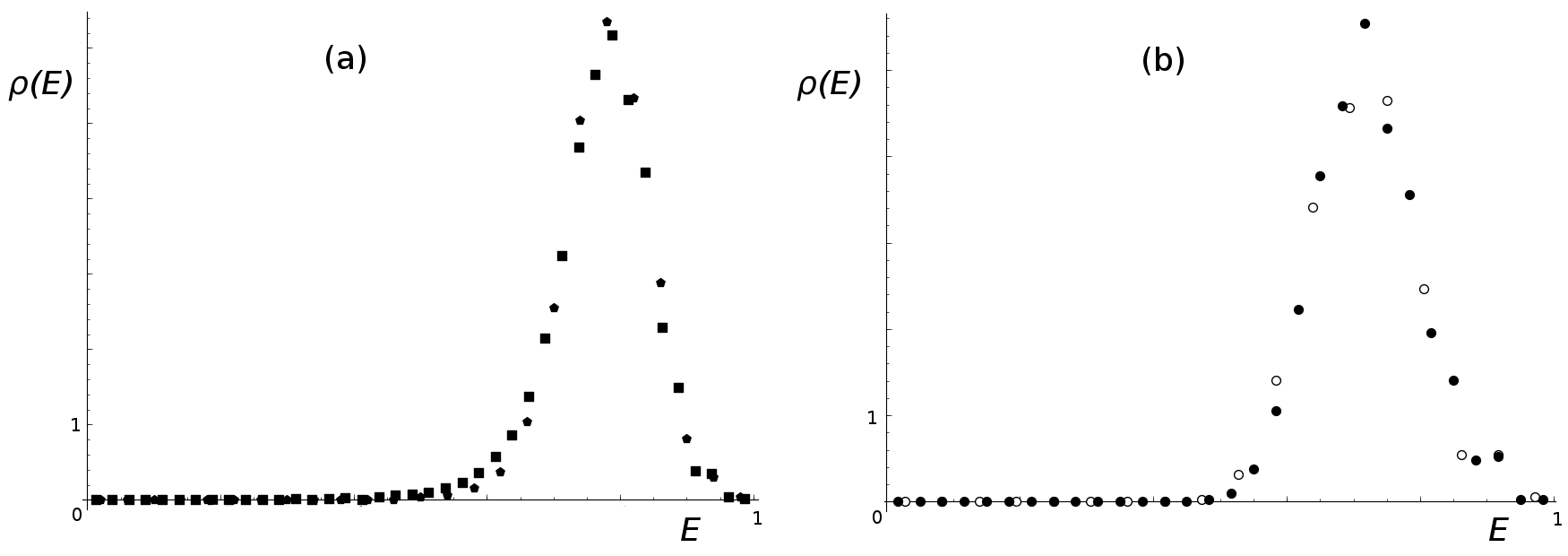}  
\caption{Distributions of normalized eigenvalues for the partially solvable cases (\ref{CMRS},\ref{CCF},\ref{CLLL}). The normalized energy $E$ is measured in units of the largest eigenvalue in each block (common for the three different resonant systems). (a) The common limiting distribution of  (\ref{CMRS}) and (\ref{CCF}): $N=M=23$ for (\ref{CCF}) corresponding to a $1255\times 1255$ matrix (pentagons) and $N=M=27$ for  (\ref{CMRS})  corresponding to a $3010\times 3010$ matrix (squares). (b) The limiting distribution for (\ref{CLLL}): $N=M=20$ corresponding to a $627\times 627$ matrix (empty circles) and $N=M=23$ corresponding to a $1255\times 1255$ matrix (filled circles).}
\label{eigsol}
\end{figure}

For the partially solvable cases (\ref{CMRS}, \ref{CCF}, \ref{CLLL}):
\begin{itemize}
\item All energy eigenvalues are non-negative,  with the maximal eigenvalue within each $(N,M)$-block given by
\beq
E_{max}^{(sol)}=\frac{N(N-1)}2,
\label{Emaxsol}
\eeq
independent of $M$, and independent of the concrete system within this class one considers. This is likely to generalize to the whole infinite family of resonant systems developed in \cite{AO}.
\item As $M$ is increased at fixed $N$, new eigenvalues are inserted between 0 and $N(N-1)/2$, while the old eigenvalues are retained. In other words, all the eigenvalues of each $(N,M_1)$-block form a subset of eigenvalues of any $(N,M_2)$-block with $M_2>M_1$. (One wonders if this `inheritance' may be explained in terms of recursive construction of partitions.)
\item Rational eigenvalues are abundant and tend to come with high multiplicities.
\item In blocks with large $N$ and $M$, distributions of normalized eigenvalues $E/E_{max}^{(sol)}$ converge to smooth bell-shaped curves, as demonstrated in Fig.~\ref{eigsol}. Note that (\ref{CMRS}) and  (\ref{CCF}) appear to converge to the same distribution, while the limiting distribution (\ref{CLLL}) looks very similar but is numerically distinct. One may imagine that these different limiting curves come from some simple family. There thus appear to be universality features in the limiting distributions within the partially solvable systems from the class introduced in \cite{AO}, though the exact details remain to be investigated. (Distributions for small $N$ and large $M$ appear to be less universal and display much more spike-like structures.)
\end{itemize}

While we have chosen to look into the distributions of eigenvalues in our above analysis, and for the special resonant systems we have considered they indeed display very attractive features asking for an analytic explanation, such eigenvalue distributions play relatively little role in quantum chaos theory \cite{haake,GMW} because they are system-specific and non-universal. On the other hand, distances between neighboring levels are considered a powerful indicator of chaotic behaviors and play a central role in quantum chaos theory. We shall therefore turn to such level spacings.

\subsection{Level spacing statistics}

A key aspect of quantum chaos theory \cite{haake,GMW} is two conjectures due to Berry-Tabor \cite{btint} and Bohigas-Giannoni-Schmit \cite{BGS} stating that distributions of properly normalized (see below) distances between neighboring energy eigenvalues for a quantum system are a good indicator of the integrable/chaotic properties of its classical limit. More specifically, the distances between energy eigenvalues for a generic integrable (in particular, not superintegrable) system are distributed as for points randomly thrown on a line, expressed by the Poisson distribution
\beq
\rho_{Poisson}(\eps)=e^{-\eps},
\label{pois}
\eeq
where $\eps$ is the properly normalized level spacing -- while level spacings of a (strongly) chaotic system are expected to obey the same statistics as for eigenvalues of an ensemble of real symmetric random matrices with independent identically distributed entries (variations of this statement are possible, say, for systems without time reversal symmetry, but that will not be important for us here). The latter distribution is very well approximated for practical purposes by the `Wigner surmise'
\beq
\rho_{Wigner}(\eps)=\frac{\pi \eps}2\, e^{-\pi\eps^2/4}.
\label{wign}
\eeq

We would  like to apply this standard lore in quantum chaos theory to the spectra within individual $(N,M)$-blocks of the resonant system (\ref{ressyst}) and see how the observed level spacing statistics correlates with the expected properties of classical dynamics for different choices of the interaction coefficients. We shall see that systems whose classical versions are known to display special analytic features immediately stand out in this analysis.

Before we can proceed, we must specify what precisely is meant by `properly normalized' level spacings. The spectral statistics conjectures are formulated for \emph{unfolded} spectra, meaning that the energy scale is locally stretched/compressed so that the mean level density within any finite energy range much bigger than individual level spacings is constant (for a contemporary discussion of unfolding procedures, see \cite{GMW,abdulmagd,luukko}). The spectral statistics thus measures `microscopic' fluctuations in energy level positions, rather than large-scale level density modulations. These fluctuations are believed to possess strong universal properties, while the large-scale density modulations are system-dependent and non-universal. The unfolding procedure is necessarily ambiguous for our finite-sized blocks (or any finite-sized samples for that matter), since one has to decide on the scale that separates the `slow modulations' that are removed from the `microscopic fluctuations' that are retained. Nonetheless, one may hope (and this is supported  by numerics) that definitions of unfolding convergent in the limit of infinite block size can be given. For our practical purposes, the following very simple-minded definition shall suffice: given a set of energy eigenvalues $E_I$ within an $(N,M)$-block with $I=1,\ldots,p_N(M)$, we choose an integer $\Delta$ close to $\sqrt{p_N(M)}$ and first define the raw unfolded sequence of level spacings
\beq
\eps_I^{(raw)}=\frac{E_{I+1}-E_{I}}{E_{I+\Delta}-E_{I-\Delta}},
\eeq
with $I=\Delta+1,\Delta+2,\ldots,p_N(M)-\Delta$. The combination $1/(E_{I+\Delta}-E_{I-\Delta})$ is proportional to the mean level density within a `smoothing' range consisting of $2\Delta$ intervals between adjacent levels and centered on the point of observation. After this raw sequence has been computed, we identify its average $\bar\eps=(\sum_{I=\Delta+1}^{p_N(M)-\Delta}\eps_i^{(raw)})/(p_N(M)-2\Delta)$, and then define the final normalized unfolded sequence
\beq
\eps_I=\frac{\eps_I^{(raw)}}{\bar\eps}.
\eeq
It is for this unfolded $\eps_I$ that the histograms are constructed and compared to (\ref{pois}) and (\ref{wign}). Note that the mean of the sequence $\eps_I$ is by definition 1, as it is for (\ref{pois}) and (\ref{wign}). There are no adjustable parameters involved in the comparison.

From the onset, we remark that, while being of great use in generic situations, the above analysis does not produce meaningful results for the Szeg\H o case (\ref{CSz}), which is highly special. The reasons are evident from the spectral features of the Szeg\H o resonant system outlined in section \ref{intpttrn}. All energy eigenvalues of the Szeg\H o system within a given $(N,M)$-block are integers between 0 and $(N-1)(N+2M)/2$, while the number of eigenvalues is $p_N(M)$ whose growth is faster than linear in $M$ unless $N=2$. Therefore, at large $M$, the spectrum is highly `overpopulated' with very large degeneracies of the integer levels, and typical distances between neighboring levels being 0 or 1, with perhaps a few other integer exceptions. The level spacing distribution thus has no chance of converging to a smooth curve at large $M$ that can be compared to (\ref{pois}) and (\ref{wign}).
All of this points to the Szeg\H o system (whose classical version has been proved Lax-integrable \cite{GG}) not being a generic integrable system (which is where the distribution (\ref{pois}) is believed to be relevant), but rather much more special. The mere observation of a perfectly integer spectrum of energies by itself alludes to the same.

\begin{figure}[t!]
\hspace{0.2cm}\includegraphics[scale=0.37]{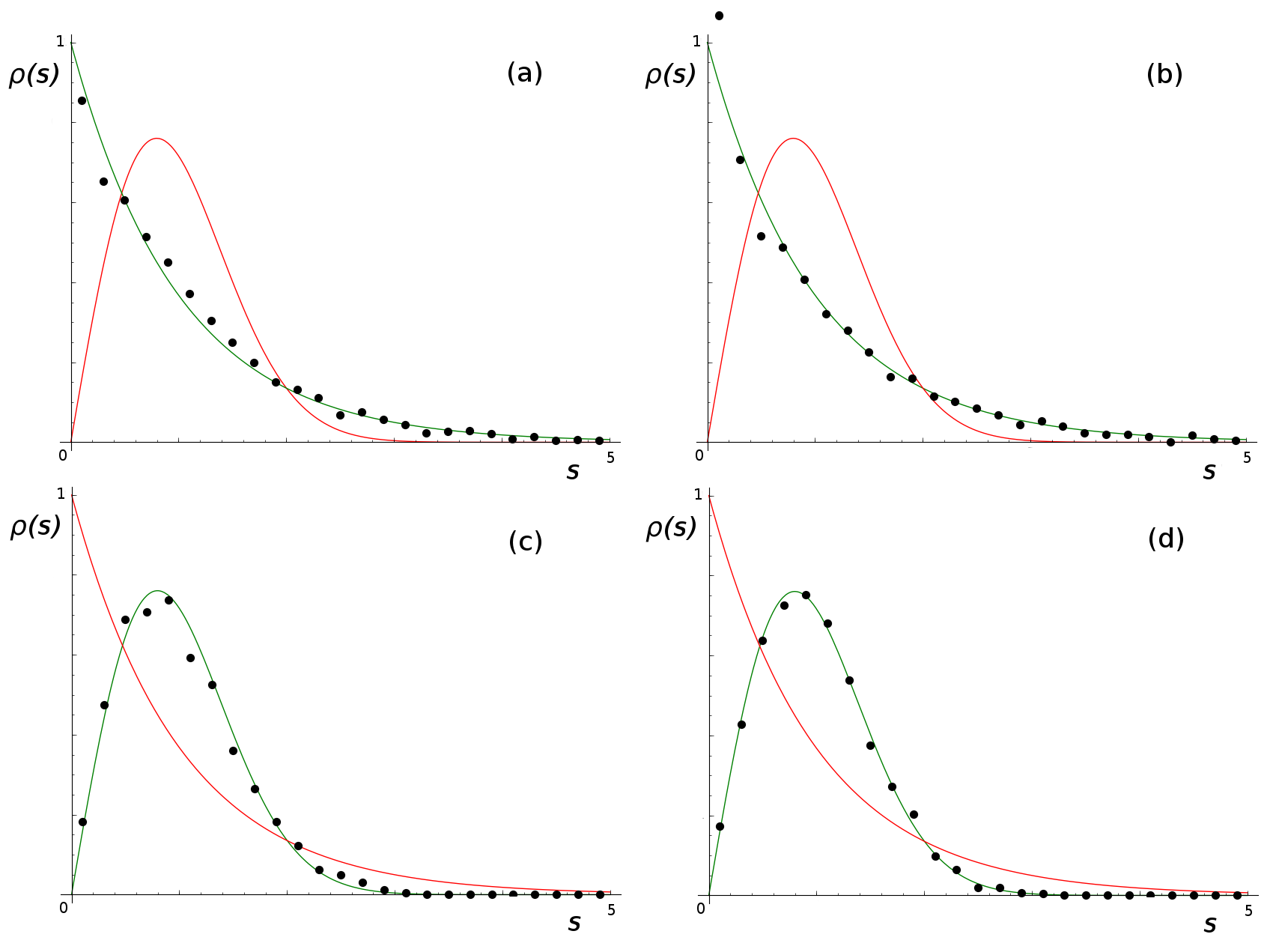}  
\caption{Unfolded level spacing distributions at $N=M=27$, compared with the Wigner surmise (the bell-shaped curve) and the Poisson statistics (the monotonically decreasing curve): (a) for the partially solvable conformal flow resonant system (\ref{CCF}) -- with the result matching the expectations for an integrable system; (b) for the partially solvable lowest Landau level resonant system (\ref{CLLL})  -- with the result matching the expectations for an integrable system; (c) for a small \emph{ad hoc} modification of the conformal flow given by (\ref{CmodCF}) -- with the result matching the expectations for a chaotic system; (d) for a randomly generated resonant system with the interaction coefficients uniformly distributed in $[0,1]$ -- with the result matching the expectations for a chaotic system.}
\label{diffdistr}
\end{figure}

We finally turn to applying the level spacing statistics methodology to the remaining systems within our pool, focusing on a specific large block with $N=M=27$ corresponding to a $3010\times 3010$ matrix. Here, our results, shown in Fig.~\ref{diffdistr}, are in a beautiful agreement with the existing quantum chaos lore. The spectral statistics clearly identifies the partially solvable systems (\ref{CCF}) and (\ref{CLLL}), defined by completely different expressions, as fundamentally similar, and assigns them to the integrable class corresponding to the distribution (\ref{pois}) -- while a minor \emph{ad hoc} modification of (\ref{CCF}) into (\ref{CmodCF}) is immediately detected as upsetting its analytic structure and assigned to the  chaotic class corresponding to the distribution (\ref{wign}), together with an arbitrary generic resonant system of the form (\ref{ressyst}), whose coefficients are assigned randomly generated values.

\section{Discussion}

We have initiated a study of quantum resonant systems (\ref{ressyst}), which form a class of remarkably simple and practically solvable interacting quantum field theories whose classical dynamics can nonetheless be very rich and complex (many such systems emerge from nonlinear wave dynamics on confined domains -- an active and largely uncharted area in contemporary PDE mathematics). We summarize below some of the  more intriguing aspects of our findings and set up a few outstanding questions.

The most striking regularities observed in our numerical experiments are evidently for the Szeg\H o case, known to be classically Lax-integrable \cite{GG} (we feel it could be appropriate to call this system after G\'erard and Grellier who introduced and extensively studied its classical version). Here, the entire spectrum consists of integers -- it is the `harmonic oscillator' of in\-ter\-acting quantum field theories! Such equidistant spectra of energy levels are remarkable and very rare (the most common examples being the one-dimensional harmonic oscillator, or multidimensional harmonic oscillators with commensurate frequencies), and the spectrum we observe is likely to point to structures beyond ordinary integrability (and in the spirit of superintegrability). Searches for (very special) mechanical systems with equidistant spectra, besides the well-known case of harmonic oscillators, have been performed in the past -- see, e.g., \cite{equid}. A famous example of an equidistant spectrum in a many-body quantum-mechanical problem is the Calogero model \cite{calogero} in a harmonic potential.\footnote{We thank Mikhail Vasiliev for drawing our attention to a few possible relations between our line of research and Calogero models.} For a very compact derivation of the spectrum, see \cite{calogero_sp}. We have furthermore observed that the distributions of eigenvalues within large $(N,N)$-blocks of the G\'erard-Grellier system appear to neatly approach the Gumbel curve, as in Fig.~\ref{eigSz}. The Gumbel curve describes the distribution of the largest value in samples  of many independent identically distributed random variables \cite{extreme}. Perhaps more relevant for us is the fact that the same curve appears in the asymptotic distribution of integer partitions (which have played a prominent role in our analysis) over different numbers of parts \cite{gumbparts}. Precise interrelations between these properties and our results remain to be explored.

For the partially integrable cases of the family described in \cite{AO}, a number of intriguing patterns emerge as well. Most obviously, the maximal eigenvalue within each Hamiltonian block is given by a simple expression (\ref{Emaxsol}) that appears to be common for the distinct cases we have studied -- it would be interesting to ascertain whether this feature is shared by the rest of the systems in the infinite class of \cite{AO}, which we consider likely. Even more suggestively, the level spacing distributions given in Fig.~\ref{diffdistr} (as well as the level spacing distributions in other blocks we have studied) classify these systems as integrable, according to the standard `quantum chaos' lore and the Berry-Tabor conjecture \cite{btint}. On the other hand, no Lax-pair structures are known for these systems classically. Could it be that the quantum case teaches us about classical integrability that has not been manifested by more conventional means? Could one use the fairly explicit quantum solution we are capable of obtaining to pin down this classical integrable structure?

Many of the features we have studied are in the spirit of random matrix theory. For instance, analyzing the eigenvalue statistics in ensembles of systems of the form (\ref{ressyst}) with random interaction coefficients is literally a random matrix problem, and a close relative of the embedded random matrix ensembles \cite{kota} (a treatment of the corresponding classical problem will appear in \cite{meloturb}). For studies of eigenvalue distributions in concrete resonant systems, such as Figs.~\ref{eigSz} and \ref{eigsol}, there is no randomness in the matrix, but for large matrices, one nonetheless observes convergence of the eigenvalue distributions to smooth simple-looking curves. One is thus dealing with eigenvalue distributions of fixed, large, highly structured matrices. Since the matrices are generated based on integer partitions, the problem also has a strong combinatorial flavor.

In our first study presented here, we have focused on the energy eigenvalues of the Hamiltonian (\ref{ressyst}). Nothing prevents one, in principle, from extending our analysis to the eigenvectors and matrix elements of the physical observables in the energy eigenbasis, which contain a wealth of physical information. In particular, key issues of quantum thermalization, such as the eigenstate thermalization hypothesis \cite{ETH}, may be effectively investigated in such a setting.

Overall, we feel that our initial study merely scratches the surface of a vast domain open for exploration. The computational simplicity and dynamical richness of the class of models we have treated here may prompt a number of possible applications, both theoretical (through providing an arena for explicit analysis of various conjectures in the field of quantum chaos) and phenomenological (say, through connections to the physics of Bose-Einstein condensates). Our setting introduces a graceful opportunity for an encounter of such diverse fields as quantum and classical chaos and integrability, turbulence studies, random matrix theory and combinatorics. One may justifiably hope that such encounter will prove fruitful.


\section*{Acknowledgments}

A number of colleagues have influenced the course of the research presented in this paper. We thank in particular
\begin{itemize}
\item Raffaele Fazio and Jorge Russo for encouraging our interest in quantizing resonant systems;
\item Alessandro Sfondrini for a useful discussion on diagonalizing lattice Hamiltonians at the beginning of this project;
\item Vijay Balasubramanian, Charles Rabideau and especially Ben Craps for discussions on quantum chaos;
\item Anxo Biasi and Piotr Bizo\'n for extensive collaboration on classical dynamics of resonant systems closely related to this work;
\item Luca Lionni for useful input on combinatorics of partitions;
\item St\'ephane Dartois, Luca Lionni, Vincent Rivasseau and Guillaume Valette for collaboration on related work \cite{meloturb} dealing with averaged classical dynamics in ensembles of resonant systems;
\item Sikarin Yoo-kong for participation in early stages of this project;
\item Maciej Nowak for a conversation on limiting distributions that encouraged our searches and eventually led to the identification of Fig.~\ref{eigSz} as the Gumbel distribition;
\item Mikhail Vasiliev for a stimulating conversation about properties of Calogero models.
\end{itemize}
The work of O.E. has been supported by CUniverse research promotion project (CUAASC) at Chulalongkorn University. W.P. is funded by Petchraprajomklao scholarship from King Mongkut's University of Technology Thonburi.


\end{document}